\documentclass[aps,prl,amsmath,amssymb,twocolumn,amsfonts,longbibliography]{revtex4-2}
\usepackage{amssymb}
\usepackage{latexsym}
\usepackage[colorlinks,allcolors=red]{hyperref}

\usepackage{epsfig}

\usepackage{dcolumn}
\usepackage{amsmath}
\usepackage{amsfonts}
\usepackage{bm}
\usepackage{color}

\newcommand{\be}{\begin{equation}}
\newcommand{\ee}{\end{equation}}
\newcommand{\bea}{\begin{eqnarray}}
\newcommand{\eea}{\end{eqnarray}}

\newcommand{\s}{\sigma}

\newcommand{\la}{\langle}
\newcommand{\ra}{\rangle}

\newcommand{\ri}{\mbox{i}}

\renewcommand{\vec}[1]{{\bm #1}}

\begin{document}
\title{A microscopic model of a fractionalized Fermi liquid}
\author{Piers Coleman}
\affiliation{
Center for Materials Theory, Department of Physics and Astronomy,
Rutgers University, 136 Frelinghuysen Rd., Piscataway, NJ 08854-8019, USA}
\affiliation{Department of Physics, Royal Holloway, University
of London, Egham, Surrey TW20 0EX, UK.}
\author{Aaditya Panigrahi}
\affiliation{Department of Physics, Cornell University, Ithaca, NY 14853, USA}
\author{Alexei Tsvelik}
\affiliation{Division of Condensed Matter Physics and Materials
Science, Brookhaven National Laboratory, Upton, NY 11973-5000, USA}

 \date{\today } 
\begin{abstract}
 In this short letter we identify a relationship between the Kondo lattice model formulated in Coleman {\it et.al}, Phys. Rev. Lett. {\bf 129}, 177601 (2022) and Ancilla Layer formulation of the Hubbard model recently proposed by  Zhang and Sachdev.  
\end{abstract}

\pacs{71.10.Hf, 71.10.Pm,71.27.+a} 

\maketitle


The concept of a fractionalized Fermi liquid (FL$^*$), in which a electron and a spin Fermi liquid co-exist, was first proposed as a way to satisfy the Oshikawa sum rule at the large-to-small  Fermi surface transition of a Kondo lattice\cite{senthil03}. Recently, this idea has re-emerged in the context of cuprate superconductors\cite{sachdev25}, to account for the identification of small Fermi-surface pockets in the underdoped compounds. In particular, a recent analysis of the angular dependence of the magnetoresistance\cite{harrison25}, which displays a Yamaji effect reveals that the area of these putative Fermi pockets  scales with the hole density, rather than the electron density of a conventional Fermi liquid. The persistence of the angle-dependent resistivity to high temperatures appears to rule out the doubling of the elementary cell due to density wave formation. These observations thus suggest the existence of a phase with Fermi surfaces that do not enclose the conventional Luttinger volume, motivating the idea that this is an FL$^*$ phase.  

Several theoretical proposals \cite{senthil03, senthil2, Ashvin, Grover, Bonderson, Tsvelik2016} have posited that an FL$^*$ phase involves the co-existence of a charged Fermi liquid and a background spin liquid which supports a sea of neutral  fermion excitations that is  invisible to  ARPES experiments. 
In this way, the missing portion of the Luttinger volume remains hidden from view. To provide a concrete formulation of the FL$^*$ concept, Zhang and Sachdev recently advanced an Ancilla Layer Model (ALM) \cite{ALM1, ALM2},  in which an ancilliary filled band decouples, via a lattice Kondo effect,  into a conduction band and a spin liquid: the former subtracts from hole Fermi surfaces, giving them an area proportional to the hole density, while the latter decouples from the Fermi surface count as the hidden component of an FL$^*$\cite{sachdev25,hazra}.  The crux of the issue depends crucially on the existence of a decoupled spin liquid of fermions. 

\begin{figure}[h]
        \includegraphics[width=1.0\linewidth]{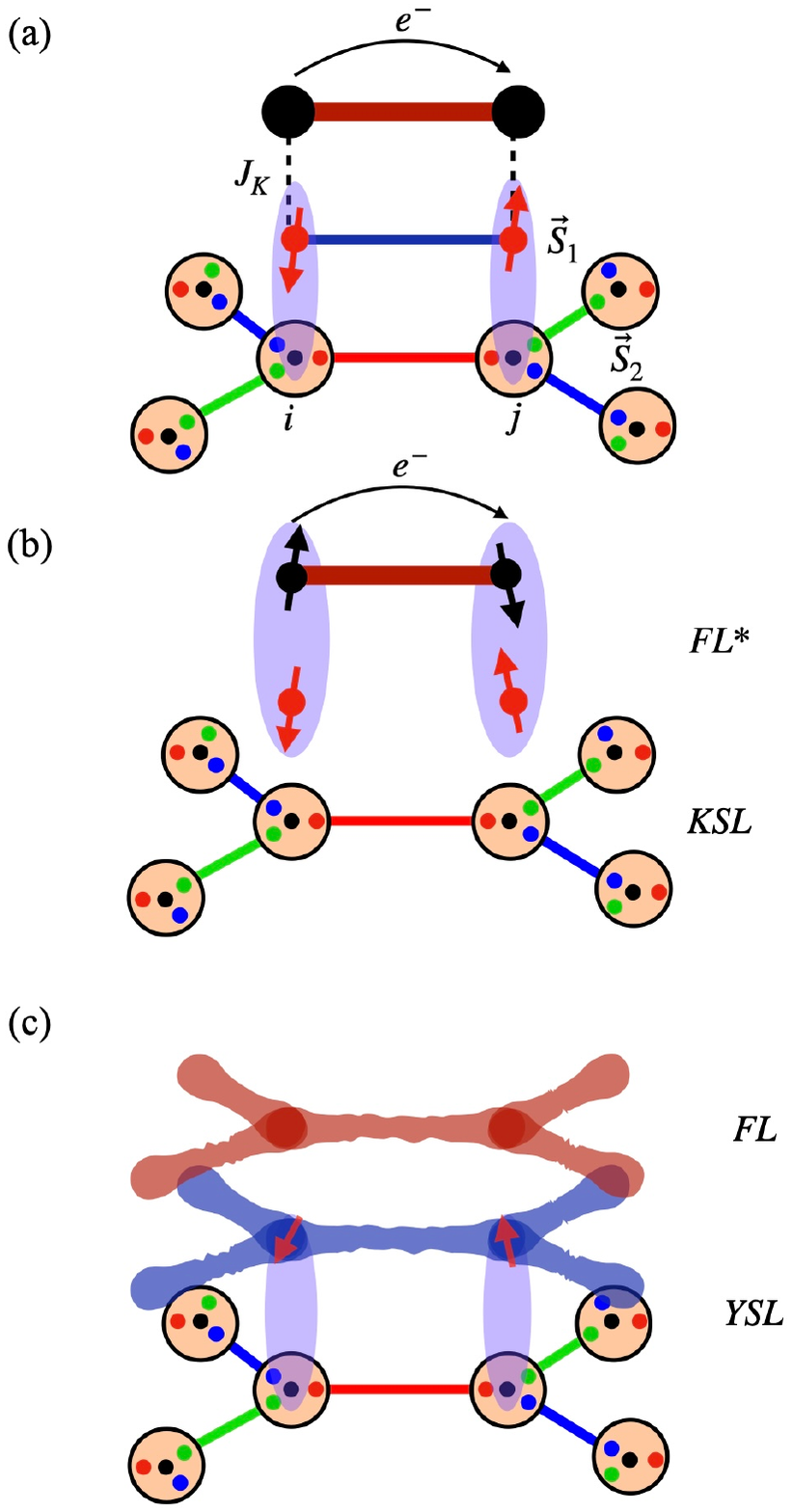}
        \caption{ a.) In the CPT model (1) each site is trivalent, hosting conduction electrons, shown here as black circles, localized spins-1/2  ${\bf S}_1$ linked by Heisenberg exchange and ${\bf S}_2$, linked by color-coded Kitaev interactions. The frustrated Kitaev interactions induce a Majorana fractionalization of spins, which are then Kondo-coupled to the conduction electrons on the same lattice. b.) FL* phase. At large Kondo coupling spins ${\bf S}_1$ are "absorbed" into the electron band contributing to its Luttinger volume. Then spins ${\bf S}_2$ create Kitaev spin liquid - a Majorana thermal metal with a Fermi surface. c.) At weak  Kondo coupling the spins do not contribute to the Luttinger volume of the conduction band. They  create Yao-Lee spin liquid instead. }   \label{fig:Fig1}
    \end{figure}

Here we  highlight a 
solvable model, the CPT model\cite{CPT1}  which  provides a concrete realization of the  FL$^*$ phase \cite{CPT2} envisioned in the ALM approach.  While there are key differences between the two approaches, the salient features have much in common. As in the ALM approach, the CPT model \cite{CPT1} can be regarded as a three-layered set of excitations, consisting of a conduction sea linked to two separate spin layers.  The middle layer is described by spin-1/2 moments with a Heisenberg symmetry, which we denote by $\mathbf{ S}_1$, while the lower layer describes second set of spin-1/2 moments,  $\mathbf{S}_2$ which couple via an anisotropic Kitaev-like Ising interactions. Here we follow the notation used in \cite{ALM1, ALM2} rather than the orbital-spin notation used in the original CPT model. 
The resulting Kondo lattice model can be regarded as a three-layer system in which 
 each site contains itinerant electrons $(c, c^+)$ and the two types of spin-1/2 operators, $\mathbf{S}_1$ and $\mathbf{S}_2$ (Fig. \ref{fig:Fig1} a).
 Both electrons and spins are situated on a hyperoctagon lattice 
which allows the dynamics of the spin fluid to be exactly solved using classic Kitaev-methods, in this case giving rise to a gapless Yao Lee spin liquid with a Fermi surface,  in isolation.  The excitations of this spin liquid are fractionalized, consisting of Majorana fermions with a Fermi surface and gapped visons - fluxes of the Z$_2$ gauge field.  The lattice couples a band of conduction electrons to the mid-layer Heisenberg spins $\mathbf{ S}_1$ of the Yao-Lee spin liquid via a Kondo interaction. The CPT Hamiltonian  thus contains three terms:
     $H_{CPT}=H_c+H_K
+H_{YL}\label{CPTHam} 
        $, where 
\begin{eqnarray}
			&& H_{c}=-t\sum_{<i,j>} ( c^+_{i\s}c_{j\s}+{\rm H.c.})-\mu\sum_{i}c^+_{i\s}c_{i\s},\cr&& H_{K}=J_K\sum_{i}(c^+_{i}\vec{\s}c_{i})\cdot \vec{S}_{1,i},\cr
			&& H_{YL}=K/2\sum_{<i,j>}S_{2,i}^{\alpha_{ij}}S_{2,j}^{\alpha_{ij}}
			(\vec{S}_{1,i}\cdot\vec{S}_{1,j}),\label{YLH}
			.
			\label{CPTComp} 
        \end{eqnarray} 
Here $H_c$ describes the nearest-neighbor hopping of the conduction electrons, $H_K$ couples the conduction sea to the middle-layer of spins $\mathbf{S}_1$ via an antiferromagnetic Kondo interaction while  $H_{YL}$ describes the Yao-Lee spin-spin interaction within the bottom two spin layers. 
The Yao-Lee term involves an anisotropic nearest-neighbor Ising interaction between the $\alpha_{ij} = x, y, z$ components of the orbitals $S_{2,j}$, which is decorated by a Heisenberg interaction between the spins $\vec{S}_{1,j}$. 

\begin{figure}[h]
        \includegraphics[width=1.0\linewidth]{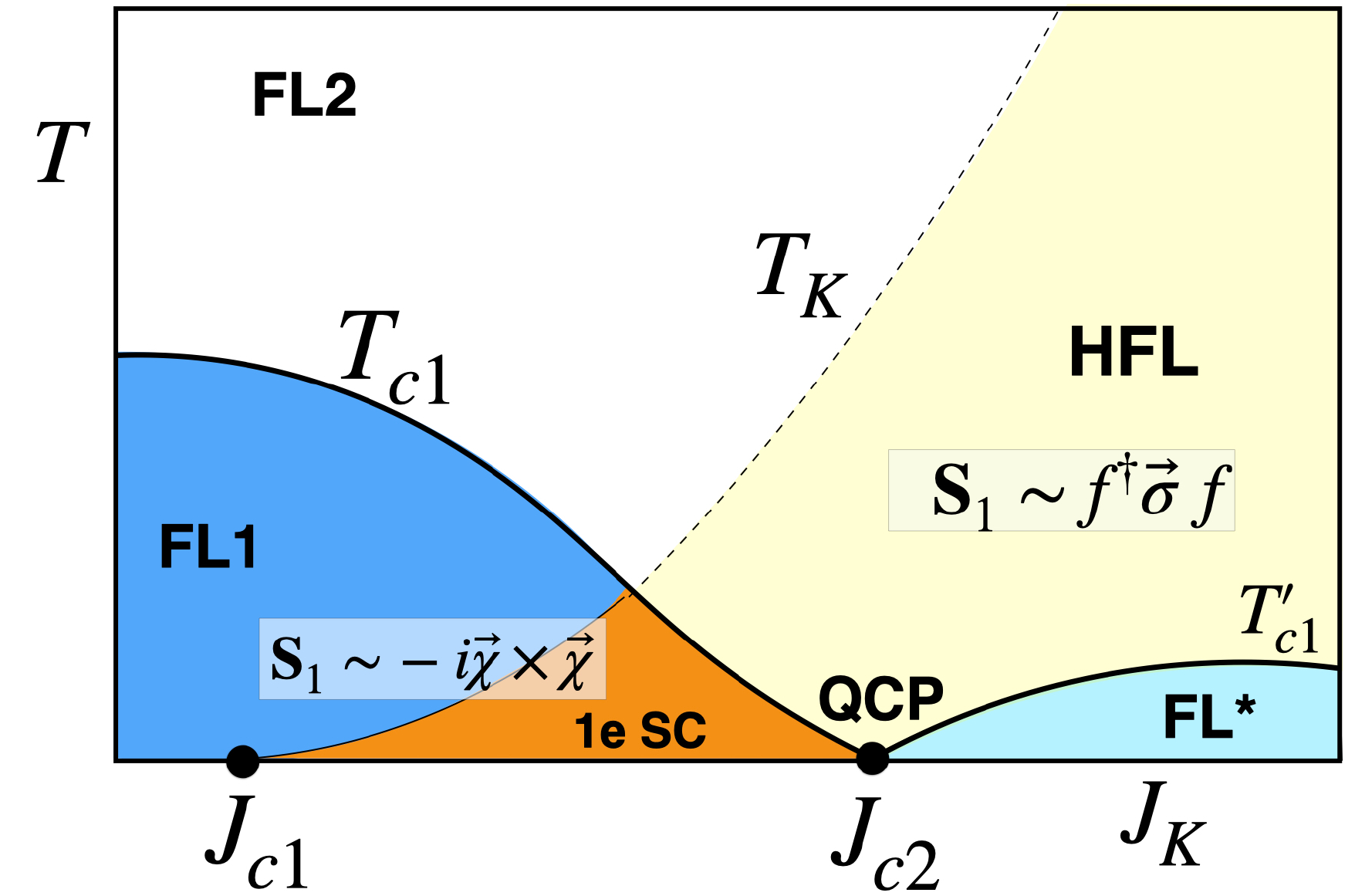}
        \caption{ Phase diagram of the CPT model Eq.(1) and Fig. 1a  for  more than one electron per site after \cite{CPT2}. In FL$_{1,2}$ phases the spin and electron subsystems are decoupled (Fig. 1c). Below the Kondo temperature $T_K$ they are coupled. At  $J_{c1} < J_K <J_{c2}$ the ground state is an exotic (fractionalized) superconductor. At $J>J_{c2}$ the ground state is an  heavy fermion liquid FL*  and fractionalized Kitaev spin liquid (Fig. 1b). The transition lines $T_{c1}, T'_{c1}$ mark phase transitions in, respectively, the Yao-Lee and Kitaev spin liquids. Below these transitions the visons become rare.} 
        \label{fig:Fig2}
    \end{figure}

.

 Although both CPT and ALM models support the FL* phase, the global phase diagrams are different. The most prominent difference is the presence of a superconducting phase separating the normal  FL and FL* phases of the CPT model.  In \cite{CPT2}, we presented the phase diagram of the CPT model at a filling corresponding to one electron per site, where the electron subsystem forms a compensated metal. We identified three distinct phases. For $J_K < 0$  (the left side of the phase diagram), the electron subsystem is decoupled from the spin liquid (SL). This phase is a conventional Fermi liquid (FL), though with a subtlety: it features two Fermi surfaces—electron and hole—of equal volume (compensated metal). There is a difference with ALM model where the decoupled spin layers form a singlet. In the FL$_1$ state the decoupled Yao-Lee spin liquid has a Majorana Fermi surface. This Fermi surface is invisible to ARPES which couples only to the electrons. 
 
 In our subsequent papers \cite{CPT3,CPT4} we discussed finite doping. This  leads to some changes in the phase diagram which, however, do not affect our main result - the existence of FL* at strong coupling. The superconducting instability now occurs at a finite threshold $J_{c1}$. The character of the superconducting phase may change depending on the doping: at stronger doping we expect  a Pair Density Wave formation. These details, however, are not relevant to the main subject of this paper which is a demonstration of the feasibility of FL* phase. This phase emerges in  the strong $J_K$ limit as a  heavy fermion metal where the Fermi surface volume includes on doping. Meanwhile the orbital  Kitaev liquid remains intact.

    Different phases of the CPT model correspond (see Fig. \ref{fig:Fig2}) to different patterns of the spin fractionalization. Spin 1/2 is unique in the sense that it can be fractionalized as a product of two Majorana fermions, as in the Kitaev model \cite{kitaev}. The tensor product of two spins  fractionalizes into bilinears of Majorana fermions $S_1^a S_2^\alpha\rightarrow - 2i \chi^a b^\alpha$, where $\chi^a$ ($a=1,3$)  and $b^\alpha$ ($\alpha = 1,3 $ are the Majorana fermions  which fractionalize $\vec S_1 = - i \vec \chi \times \vec \chi $ and $S_2= -i \vec b \times \vec b$ respectively. It is this  formed from the six Majoranas, and allows to fermionize the Yao-Lee model \cite{YL}. This fractionalization pattern works in the exotic fractionalized superconducting phase which exists in the CPT model at small $J_K$ \cite{CPT1}. The FL* phase appears at strong positive $J_K $ (the right side of the phase diagram Fig. 2). The corresponding fractionalization pattern here is more conventional:
  \bea
  {\bf S}_1 = \frac{1}{2}f^+_{\alpha}\vec\s_{\alpha\beta}f_{\beta}, ~~ \sum_{\alpha}f^+_{\alpha}f_{\alpha} =1,
  \eea
  which allows one to decouple the interaction as
  \bea
  J_K\sum_{i}(c^+_{i}\vec{\s}c_{i})\cdot \vec{S}_{1,i} \rightarrow \Phi^2/2J_K + \Phi (c^+f + f^+c),
  \eea
  (Ref. \cite{ALM2} uses an alternative representation where the spinon operators are represented by Nambu spinors and the corresponding Higgs field is a complex 2$\times$2 matrix). In the saddle point approximation the Higgs field $\Phi$ acquires a finite expectation value $\Phi \sim t\exp[-1/\rho(\epsilon_F)J_K]$ and  both itinerant electrons and holes hybridize with the $f$-operators, that is with ($\mathbf{S}_1$) spins.  At one electron per site this  results in a Kondo insulator—effectively two Kondo insulators, one for electrons and one for holes  since at one electron per site the electron Fermi surface consists of electron and hole pockets. At finite doping we get a metal, the $\mathrm{FL}^*$. 
By projecting the Yao-Lee Hamiltonian (\ref{YLH}) onto the ground state of this Kondo insulator/heavy fermion metal, we obtain the Kitaev spin liquid for the ($\mathbf{S}_2$) spins:

  \bea
  && \la 0_{KI}|H_{YL}|0_{KI}\ra =K/2\sum_{<i,j>}S_{2,i}^{\alpha_{ij}}S_{2,j}^{\alpha_{ij}}
			\la (\vec{S}_{1,i}\cdot\vec{S}_{1,j})\ra = \nonumber\\
            && K^*\sum_{<i,j>}S_{2,i}^{\alpha_{ij}}S_{2,j}^{\alpha_{ij}} = \ri K^*\sum_{<i,j>}u_{ij}\chi_i\chi_j,
  \eea
  where $u_{ij} = \pm 1$ is the Z$_2$ static gauge field and $\chi_i$ are Majorana fermions. As we have noticed above, on the hyperoctagonal lattice the Majorana fermions of the  Kitaev spin liquid have  a Fermi surface. As is mentioned above, a distinct feature of our model is the intermediate phase between FL and FL* - the fractional superconductor. This feature is not universal; its existence is guaranteed only in the presence of approximate nesting between the electron and Majorana FS's. 


      In \cite{Square} we generalized the above construction for square lattice. The price to pay was an increased number of degrees of freedom.  Namely, the corresponding Yao-Lee model includes not two spins 1/2 on one site, but three. The resulting  spin liquid can be formulated as a theory of pseudospin 1/2 Dirac fermions.  

   {\it Conclusions.} This paper has emphasized how a  Yao-Lee spin liquid embedded within a Kondo lattice, the CPT model,  provides a tractable model of a fractionalized Fermi liquid or FL$^*$, establishing an interesting connection to the Ancillar Layer Model(ALM) approach to the Hubbard model.  The $Z_2$ character of the FL$^*$ in this approach may be an interesting feature to examine in the context of the ALM approach. 
   

 {\it Acknowledgments. }
 AT is grateful to A. Weichselbaum and W. Yin for valuable discussions. This work was supported by Office of Basic Energy Sciences, Material
		Sciences and Engineering Division, U.S. Department of Energy (DOE)
		under Contracts No. DE-SC0012704 (AMT) and DE-FG02-99ER45790 (PC ). AP was supported by the U.S. Department of Energy, Office of Science, Basic Energy Sciences, Materials Sciences and Engineering Division

\bibliography{flstar.bib}

@article{CPT1, title={Solvable 3D Kondo Lattice Exhibiting Pair Density Wave, Odd-Frequency Pairing, and Order Fractionalization}, volume={129}, ISSN={0031-9007}, DOI={10.1103/physrevlett.129.177601}, abstractNote={The Kondo lattice model plays a key role in our understanding of quantum materials, but a lack of small parameters has posed a long-standing problem. We present a three-dimensional S=1/2 Kondo lattice model describing a spin liquid within an electron sea. Strong correlations in the spin liquid are treated exactly, enabling a controlled analytical approach. Like a Peierls or BCS phase, a logarithmically divergent susceptibility leads to an instability into a new phase at arbitrarily small Kondo coupling. Our solution captures a plethora of emergent phenomena, including odd-frequency pairing, pair density wave formation and order fractionalization. The ground-state state is a pair density wave with a fractionalized charge e, S=1/2 order parameter, formed between electrons and Majorana fermions.}, number={17}, journal={Physical Review Letters}, author={Coleman, Piers and Panigrahi, Aaditya and Tsvelik, Alexei}, year={2022}, pages={177601} }

@article{CPT2, title={Breakdown of order fractionalization in the CPT model}, volume={110}, ISSN={2469-9950}, DOI={10.1103/physrevb.110.104520}, abstractNote={We present an analysis of the half-filled CPT model, an analytically tractable Kondo lattice model with Yao-Lee spin-spin interactions on a 3D hyperoctagon lattice, proposed by Coleman, Panigrahi, and Tsvelik. Previous studies have established that the CPT model exhibits odd-frequency triplet superconductivity and order fractionalization. Through asymptotic analyses in the small-J and large-J Kondo coupling limits, we identify a quantum critical point at Jc, marking a transition from a superconductor to a Kondo insulator. By estimating the vison gap energy to account for thermal gauge fluctuations, we determine the energy scales governing the thermal breakdown of order fractionalization. Moreover, at large J the Kondo insulator undergoes orbital decoupling, leading to the formation of a decoupled Kitaev orbital liquid. These findings and analogies with the Z2-gauged XY model lead us to propose a tentative phase diagram for the CPT model at half-filling.}, number={10}, journal={Physical Review B}, author={Panigrahi, Aaditya and Tsvelik, Alexei and Coleman, Piers}, year={2024}, pages={104520} }

@article{senthil03,
  author = {T. Senthil and S. Sachdev and M. Vojta},
  title = {Fractionalized Fermi Liquids},
  journal = {Physical Review Letters},
  volume = {90},
  pages = {216403},
  year = {2003},
  doi = {10.1103/PhysRevLett.90.216403}
}

@article{CPT3, title={Microscopic Theory of Pair Density Waves in Spin-Orbit Coupled Kondo Lattice.}, volume={135}, ISSN={0031-9007}, DOI={10.1103/pdqz-zb8k}, abstractNote={We demonstrate that the discommensuration between the Fermi surfaces of a conduction sea and an underlying spin liquid provides a natural mechanism for the spontaneous formation of pair density waves. Using a recent formulation of the Kondo lattice model that incorporates a Yao Lee spin liquid proposed by the authors, we demonstrate that doping away from half filling induces finite-momentum electron-Majorana pair condensation, resulting in amplitude-modulated pair density waves (PDWs). Our approach provides a precise, analytically tractable pathway for understanding the spontaneous formation of PDWs in higher dimensions and offers a natural mechanism for PDW formation in the absence of Zeeman splitting.}, number={4}, journal={Physical review letters}, author={Panigrahi, Aaditya and Tsvelik, Alexei and Coleman, Piers}, year={2025}, pages={046504} }

@article{sachdev25,
  author = {P. M. Bonetti and M. Christos and A. Nikolaenko and A. A. Patel and S. Sachdev},
  title = {Critical quantum liquids and the cuprate high temperature superconductors},
  journal = {arXiv preprint},
  eprint = {2508.20164},
  year = {2025}
}

@article{harrison25,
  author = {M. K. Chan and K. A. Schreiber and O. E. Ayala-Valenzuela and E. D. Bauer and A. Shekhter and N. Harrison},
  title = {Observation of the Yamaji effect in a cuprate superconductor},
  journal = {Nature Physics},
volume ={21},
pages ={1753},
  year = {2025},
doi ={10.1038/s41567-025-03032-2}
}

@article{senthil2,
  author = {T. Senthil and M. Vojta and S. Sachdev},
  title = {Weak magnetism and non-Fermi liquids near heavy-fermion critical points},
  journal = {Physical Review B},
  volume = {69},
  pages = {035111},
  year = {2004},
doi ={10.1103/PhysRevB.69,035111},
  eprint = {cond-mat/0305193}
}

@article{Ashvin,
  author = {A. Paramekanti and A. Vishwanath},
  title = {Extending Luttinger’s theorem to Z2 fractionalized phases of matter},
  journal = {Physical Review B},
  volume = {70},
  pages = {245118},
  year = {2004},
doi ={10.1103/PhysRevB.70.245118},
  eprint = {cond-mat/0406619}
}

@article{Grover,
  author = {T. Grover and T. Senthil},
  title = {Quantum phase transition from an antiferromagnet to a spin liquid in a metal},
  journal = {Physical Review B},
  volume = {81},
  pages = {205102},
  year = {2010},
doi ={10.1103/PhysRevB.81.205102},
  eprint = {0910.1277}
}

@article{Bonderson,
  author = {P. Bonderson and M. Cheng and K. Patel and E. Plamadeala},
  title = {Topological Enrichment of Luttinger’s Theorem},
  journal = {arXiv e-prints},
  year = {2016},
  eprint = {1601.07902}
}

@article{Tsvelik2016,
  author = {A. M. Tsvelik},
  title = {Fractionalized Fermi liquid in a Kondo-Heisenberg model},
  journal = {Physical Review B},
  volume = {94},
  pages = {165114},
  year = {2016},
doi ={10.1103/PhysRevB.94.165114},
  eprint = {1604.06417}
}

@article{hazra,
  author = {T. Hazra and P. Coleman},
  title = {Luttinger sum rules and spin fractionalization in the SU(N) Kondo Lattice},
  journal = {Physical Review Research},
  volume = {3},
  pages = {033284},
  year = {2021},
doi ={10.1103/PhysRevResearch.3.033284},
}

@article{ALM1,
  author = {Y.-H. Zhang and S. Sachdev},
  title = {From the pseudogap metal to the Fermi liquid using ancilla qubits},
  journal = {Physical Review Research},
  volume = {2},
  pages = {023172},
  year = {2020},
doi={10.1103/PhysRevResearch.2.023172},
  eprint = {2001.09159}
}

@article{ALM2,
  author = {Y.-H. Zhang and S. Sachdev},
  title ={Deconfined criticality and ghost Fermi surfaces at the onset of
antiferromagnetism in a metal}, 
journal ={Phys. Rev. B},
volume ={102},
pages ={155124},
year = {2020},
eprint ={2006.01140},
doi={10.1103/PhysRevB.102.155124}
}

@article{kitaev,
author = {Kitaev, Alexei},
title= {Anyons in an exactly solved model and beyond},
journal ={Annals of Physics},
volume ={321},
pages ={2 –111},
year ={2006}
}

@article{YL,
    author ={Hong Yao and Dung-Hai Lee},
title ={Fermionic magnons, non-abelian
spinons, and the spin quantum Hall effect from an exactly solvable spin-1/2 Kitaev model with su(2) symmetry},
journal ={Phys. Rev.Lett},
volume ={107},
pages ={087205},
year ={2011},
doi={10.1103/PhysRevLett.107.087205}
}

@article{Square,
  author = {Piers Coleman and Elio  K\"onig and Aaditya Panigrahi  and Alexei Tsvelik},
  title = {Tractable model for a fractionalized Fermi liquid (FL*) on a square lattice},
  journal = {arXiv e-prints},
  year = {2026},
  eprint = {2604.06157}
}

@article{CPT4,
  author = {Piers Coleman and Aaditya Panigrahi  and Alexei Tsvelik},
  title = { Bose metal near pair-density-wave order in a spin-orbit-coupled Kondo lattice},
  journal = {arXiv e-prints},
  year = {2026},
  eprint = {2604.184501}
}
\end{document}